\documentclass[prd,twocolumn,nofootinbib,tightenlines,floatfix,showpacs ,amssymb, aps,superscriptaddress]{revtex4-1}
\usepackage{graphicx}
\usepackage{dcolumn}
\usepackage{bm}
\usepackage{epsfig}
\usepackage{amsmath}
\usepackage{xcolor}
\begin{document}

\title{Fast Radio Bursts From Primordial Black Hole Binaries Coalescence}

\author{Can-Min Deng}
\email{dengcm@pmo.ac.cn}
\affiliation{Purple Mountain Observatory, Chinese Academy of Sciences, Nanjing 210008, China}
\affiliation{School of Astronomy and Space Science, University of Science and Technology of China, Hefei, Anhui 230026, China}

\author{Yifu Cai}
\email{yifucai@ustc.edu.cn}
\affiliation{School of Astronomy and Space Science, University of Science and Technology of China, Hefei, Anhui 230026, China}
\affiliation{CAS Key Laboratory for Researches in Galaxies and Cosmology, Department of Astronomy, University of Science and Technology of China, Hefei, Anhui 230026, China}

\author{Xue-Feng Wu}
\email{xfwu@pmo.ac.cn}
\affiliation{Purple Mountain Observatory, Chinese Academy of Sciences, Nanjing 210008, China}
\affiliation{School of Astronomy and Space Science, University of Science and Technology of China, Hefei, Anhui 230026, China}

\author{En-Wei Liang}
\email{lew@gxu.edu.cn}
\affiliation{GXU-NAOC Center for Astrophysics and Space Sciences, Department of Physics, Guangxi University, Nanning 530004, China}
\affiliation{Guangxi Key Laboratory for the Relativistic Astrophysics, Nanning 530004, China}

\begin{abstract}
In this paper we propose the model that the coalescence of primordial black holes (PBHs) binaries with equal mass $M \sim 10^{28}$g can emit luminous gigahertz (GHz) radio transient, which may be candidate sources for the observed fast radio bursts (FRBs), if at least one black hole holds appropriate amount of net electric charge $Q$. Using a dimensionless quantity for the charge $q = Q/\sqrt{G}M$, our analyses infer that $q\sim O(10^{-4.5})$ can explain the FRBs with released energy of order $O(10^{40}) {\rm ergs}$. With the current sample of FRBs and assuming a distribution of charge $\phi(q)$ for all PBHs, we can deduce that its form is proportional to $q^{-3.0\pm0.1}$ for $q\geq 7.2\times10^{-5}$ if PBHs are sources of the observed FRBs. Furthermore, with the proposed hypothetical scenario and by estimating the local event rate of FRBs $\sim 2.6 \times 10^3 {\rm Gpc}^{-3} {\rm yr}^{-1}$, one derives a lower bound for the fraction of PBHs (at the mass of $10^{28}$g) against that of matter $f_{\rm PBH}(10^{28}{\rm g})$ $\gtrsim 10^{-5}$ needed to explain the rate. With this inspiring estimate, we expect that future observations of FRBs can help to falsify their physical origins from the PBH binaries coalescences. In the future, the gravitational waves produced by mergers of small black holes can be detected by high frequency gravitational wave detectors. We believe that this work would be a useful addition to the current literature on multimessenger astronomy and cosmology.

\end{abstract}

\maketitle

\section{Introduction}

Primordial black holes (PBHs), thought to be produced in the extremely early universe \cite{Hawking 1971, Carr 1974, Carr 1994, Zel'dovich 1967}, remain mysterious to cosmologists. As was pointed out in Refs.~\cite{Zel'dovich 1967, Hawking 1971}, they can be formed if there were over-dense cosmological fluctuations in the primordial universe. Accordingly, many theoretical mechanisms were put forward to generate PBHs and thus our understanding on the theory of cosmological perturbations has been greatly enriched thanks to the extensive study on this topic (e.g., see \cite{Garcia-Bellido:2017mdw, Domcke:2017fix, Kannike:2017bxn, Carr:2017edp, Cotner:2016cvr, Ballesteros:2017fsr, Hertzberg:2017dkh, Franciolini:2018vbk, Chen:2016kjx, Quintin:2016qro, Cai:2018tuh} for various model realizations and see \cite{Khlopov:2008qy, Sasaki:2018dmp} for comprehensive reviews).
Depending on the underlying theoretical mechanisms, these PBHs could possess different properties that would lead to different observational interests, namely, PBHs can be viewed as promising candidates for dark matter (DM) \cite{Khlopov:2008qy, Frampton 2010, Carr 2016}. In general, however, they share a common property, i.e., their abundance can be constrained by a variety of astronomical experiments relying on the mass \cite{Carr 2016}. Namely, the PBH abundance with mass below $10^{15}{\rm g}$ is strongly constrained by extragalactic $\gamma$-rays background from evaporation, and that with mass above $10^{36}{\rm g}$ is tightly constrained by cosmic microwave background observation, but so far the experimental limit on parameter space associated with mass varying from $10^{16}{\rm g}$ to $10^{35}{\rm g}$ is relatively loose. In this regard, it is of observational interest to impose further constraint on PBHs by the accumulated high precision astronomical instruments \cite{Sasaki:2018dmp}.

On the other hand, fast radio bursts (FRBs) are bright radio transient events at GHz frequencies whose physical origin still remain unknown \cite{Lorimer 2007}. The events of FRBs are characterized as having a pulsewidth of the millisecond scale \cite{Lorimer 2007,Thornton 2013, Keane 2016}. All these events show extra-galactic origins and their released energy flux span on several orders of magnitude $10^{39}$-$10^{42}{\rm erg}$ \cite{Deng 2018}. Moreover, most of FRBs are non-repeating except for FRB 121102 \cite{Spitler 2014, Spitler 2016}. In the literature there were many attempts to explain the sources of these FRB signals (e.g., see \cite{Wang 2016, Dai 2016, Zhang 2017, Ye:2017lqn} and references there in). However, so far there is no way to address all properties of FRBs including the possibly repeating behaviour, GHz radiation bands, high energy fluxes, short durations and high event rates. This issue may imply that the sources of FRBs could have different physical origins \cite{Palaniswamy 2018}.

In this paper we connect the aforementioned two mysteries together in a uniform astrophysical paradigm. It was addressed that PBHs could form binaries and coalesce within the age of the universe \cite{Nakamura 1997, Bird 2016}, further investigated by \cite{Sasaki 2016, Ali 2017}. Let us hypothetically assume at least one of the two merging PBHs holds a certain amount of charge. Einstein-Maxwell theory predicts that such a binary can radiate electromagnetic (EM) energy \cite{Zilh 2012, Zilh 2014, Liebling 2016}. Accordingly, it is possible for us to `see' PBHs by detecting such EM signals. It is interesting to note that, the paradigm of the charged BH merger to generate a brief EM counterpart has been proposed in \cite{Zhang 2016}, which mainly applied the magnetic dipolar radiation power.  Also, \cite{Liu 2016} proposed FRBs can originate in the collapse of the magnetospheres of Kerr-Newman black holes. We suggest that some non-repeating FRBs might be interpreted as the direct radio emission of charged PBHs binary coalescence.
Hence, it is reasonable to explore a variety of implications of FRB-detection to set bounds on the abundance and the charge of PBHs. As is well-known,  the exterior space-time around black holes (BHs) is uniquely determined by three, so called, `hairs', including mass $M$, angular momentum $J$, and charge $Q$ \cite{Wald 1972}. Numerous BHs have now been identified through electromagnetic and gravitational wave observations, and their masses are measured, furthermore, some of them whose spins are also estimated \cite{Webster 1972, Bolton 1972, Casares 2014, Reynolds 2014, Abbott 20161}. However, charged black holes have not been reported in the literature \cite{Zaumen 1974}. In principle, BHs could be charged when they are born or by evolution after born. For instance, a charged PBH could form in a region where there is an unequal number of charged particles \cite{Hawking 1971}. Astronomically, electrically charged compact stars could collapse to form charged BHs \cite{Ray 2003}. Once BHs get charged, Ref. \cite{Punsly 2001} points out that certain stable structures can protect BH from electric discharge. In addition, BHs could also be charged by means of interacting with environmental matter fields \cite{Wald 1974, Suffern 1975, Lee 2001, Zhang 2016}. Accordingly, we assume that the PBHs' population satisfies a charge distribution, which could be a function of the charge parameter itself and is naturally cut off by the PBH mass due to the requirement of theoretical consistency.

\section{The Model}
For simplicity, we consider a toy model of a BH binary system with non-spinning equal point mass ($M_{\rm BH}$) and one of the BHs has a larger electric charge ($Q$). We ignore the charges of the other one and treat it as neutral. It can be said the charged BH acts as a giant charged particle in this model. We will calculate the energy of electromagnetic radiation released during merging under the Newtonian approximation, and then, compare it with the numerical estimation within the framework of general relativity (GR).  We note that the electric dipole moment of the BH system is $Q\vec{r}$, where $\vec{r}$ is the position vector of the charged BH with respect to the the center of mass of the binary. The electric dipole moment of the system varies quasi-periodically with inspiral motion of the charged BH. As we known in classical electrodynamics, the EM radiation from a system of charges can be decomposed into multipolar terms. The leading-order term is electric dipole. Thus, the EM radiation luminosity from the system  is dominated by the dipolar radiation. Then the radiation power of the BH binary with the dipole approximation is given by, $P = {2Q^{2}|\ddot{\vec{r}}|^{2}}/{(3c^{3})}$ \cite{Jackson 1998}, where $c$ is the speed of light.  By taking the limit of Newtonian approximation $|\ddot{\vec{r}}_{N}|$ and including the shrinkage of the orbit due to gravitational radiation (GW) $|\ddot{r}_{GW}|$ effect, with small charge, one gets
\begin{equation} \label{eq-r}
|\ddot{\vec{r}}| \simeq |\ddot{\vec{r}}_{N}| + |\ddot{r}_{GW}| ,
\end{equation}
where $|\ddot{\vec{r}}_{N}|= GM_{\rm BH}/d^{2}$, $d=2r$ being the separation distance between two BHs.
Due to gravitational radiation, the separation distance between two BHs evolve as (for equal mass and ellipticity $e=0$) \cite{Peters 1963}
\begin{equation} \label{eq-ddt}
\dot d{\rm{ = }} - \frac{{2c}}{5}{l^{ - 3}} ~,
\end{equation}
here we introduce a dimensionless distance $l \equiv d/2r_{s}$ with $r_{s}=2GM_{\rm BH}/c^{2}$ being the Schwarzschild radius of the BH.
Accordingly, this implies
\begin{equation} \label{eq-P}
P \simeq \frac{1}{{384}}\frac{{{c^5}}}{G}{q^2}{l^{ - 4}}\left( {1 + \frac{{576}}{{625}}\,{l^{ - 10}}} \right) ~,
\end{equation}
here we have also introduced a dimensionless charge quantity $q\equiv Q/\sqrt{G}M_{\rm BH}$ which satisfies the limit $q\ll1$.
The second term in the brackets of Eq.\eqref{eq-P} is introduced by the shrinkage of the orbit. This term becomes important when the black hole rapidly spins, because $l < 1$ at the merger. However, it must be noted that $|\dot d| > c$ for $l < 0.73$ according to Eq.\eqref{eq-ddt}. Therefore, Eq.\eqref{eq-ddt} overestimates the shrinking rate of the orbit, thus overestimating $\ddot{r}_{GW}$, when $l \to 0.73$. In this model, we consider the case of non-spinning black holes and satisfy $l \ge 1$ during merger. Consequently, the second term in the brackets of Eq.\eqref{eq-P} can be neglected for orders of estimation. Hereafter, we adopt this approximation.

Note that, in this toy model, $l=1$ corresponds to the end of the BH coalescence\cite{Buonanno 1999}. From somewhere $l$ to the end of coalescence, integration of the radiating power in $l$ yields an estimate of the total radiated energies to be,
\begin{equation} \label{eq-E}
E \simeq \int_{l}^{1} d l ~ \frac{P}{ \dot{l} } =\frac{5}{6}q^{2}M_{\rm BH}c^{2}\ln l ~,
\end{equation}
It deserves to mention that, the energy radiated by charged BH binary coalescence was numerically calculated for the case of $l=2$ in \cite{Liebling 2016} where the released energy follows $E \simeq 0.9 q^{2} M_{\rm BH} c^{2}$ for the same case discussed in the present study. By adopting $l=2$ in Eq. \eqref{eq-E}, our estimate yields $E \simeq 0.6q^{2}M_{\rm BH}c^{2}$, which is consistent with the numerical result of \cite{Liebling 2016} at the same order of magnitude.

Moreover, one can estimate the timescale for coalescence by integrating Eq. \eqref{eq-ddt}, which gives
\begin{equation} \label{eq-tau}
\tau= \frac{5GM_{\rm BH}}{2c^{3}}(l^{4}-1) ~.
\end{equation}

Then, we analyze the spectrum of radiation emitted by the BH binary coalescence. In the context of electrodynamics, the frequency of EM radiation of an electric-dipole oscillator is just its oscillation frequency. With the Kepler formula $d=(2GM_{\rm BH}/\omega^{2})^{1/3}$ and Eq. \eqref{eq-P}, one gets a intrinsic  spectrum power $dP/d\omega \propto \omega^{5/3}$. This is consistent with the numerical result in \cite{Liebling 2016}. However, observationally, one actually obtains the time-integrated spectrum. We also can obtain the time-integrated spectrum $dE/d\omega  \propto {\omega ^{ - 1}}$  according to Eq.(4) and the Kepler formula. Due to the fact that two BHs would eventually merge together, the spectrum would be exponentially cut off at $\omega_{\rm{p}} \sim \frac{{c/2 \sqrt 2 }}{{{r_s}}}$. Accordingly, we have the peak frequency in our model
\begin{equation} \label{eq-omega-p}
\omega _{\rm{p}} \simeq \frac{c^{3}}{4\sqrt{2}GM_{\rm BH}} ~.
\end{equation}
This is consistent with the numerical result at the same order of magnitude. Specifically, the detailed numerical computation shows that the exponential cutoff occurs around the half of ringdown oscillation frequency $\omega_{\rm{p}} \approx c^{3}/(5GM_{\rm BH})$ \cite{Liebling 2016,Ruffini 1972, Berti 2009, Shibata 2016}.

The above analysis provides a good approximation to the numerical calculation of GR. Therefore, if the masses of PBHs are of appropriate values, the corresponding electromagnetic signals can serve as possible sources of non-repeating FRBs.
As mentioned at the beginning of this paper, FRBs detected in the GHz band have very short durations. Their spectrum generally are power law in narrow bandpass with various indices. Also, their all sky event rate can be as high as $\sim 10^{4} {\rm day}^{-1} {\rm sky}^{-1} $ \cite{Deng 2018}. To distinguish the FRBs produce by the other models from our model, we will show that FRBs produced by our model have distinct features. As mentioned above, the radiation is dominated by electric dipole component. Actually, the system also have non-zero electric quadrupole moment $|\int{(3{{x}_{i}}{{x}_{j}}-{{r}^{2}}{{\delta }_{ij}})}\rho (\vec{x}){{d}^{3}}\vec{x}|\simeq 2Q{{r}^{2}}$ \cite{Jackson 1998}. Therefore, the second-leading-order of radiation is electric quadrupole. Accordingly, we can estimate that the power emitted in electric-quadrupole radiation is smaller than the power emitted in electric-dipole radiation by a factor of $(15l)^{-1}$ \cite{Jackson 1998}. This dipole-quadrupole feature in FRBs radiation shall be the unique theoretical prediction that might be justified by future FRB experiments.

The PBHs could have produced in the early universe due to large initial density fluctuations with a wide range of masses. Their masses are
approximately comparable with the particle horizon mass at the time of PBH formation, namely, $M_{\rm PBH}\sim10^{28} \times (t/10^{-10} {\rm s}) {\rm g}$ \cite{Hawking 1971, Carr 2016}, where $t$ is estimated as the formation time of PBHs. The smallest unit of time is the Planck time $t_{P}=(\hbar G/c^{5})^{1/2} \sim 10^{-43}{\rm s}$, therefore, the mass of PBHs could be $10^{-5}$g upwards. In principle, our model can be applied to BH binaries with arbitrary mass. However, on the one hand, the merger rate of PBH binaries are proportion to $M_{PBH}^{-32/37}$\cite{Ali 2017} (also seen in the next section), which means the event rate of mergers are much more lager for PBH binaries with smaller masses. On the other hand, the electric-diploe radiation during merging is lower frequency radiation for PBH binaries with larger masses, such as stellar-mass PBH binaries. It is not clear that whether the energy in low frequency radiation can convert to {\rm GHz} signals observed as FRBs effectively. \cite{Zhang 2016} proposed that this conversion may be achieved by the magnetosphere around BH similar to the case of radio pulsars. In contrast, the electric dipole radiation from small-mass PBH binaries can  be located in high frequency band. For the two reasons above, in this work, we concentrate on the PBH binaries with  masses which electric-diploe radiation is within the GHz band, in which the radiation mechanism is straightforward. From Eq. \eqref{eq-omega-p}, $\nu_{p} = \omega_{p}/2 \pi \sim (10^{28} {\rm g} / M_{\rm PBH}) \times 1~ {\rm GHz}$. If the masses of PBHs are of order $O(10^{28})$g, the electric-diploe radiation from the PBH binary coalescence are exactly within the GHz band and hence behave as fast radio bursts. The kinetic energy of the orbiting PBH can be estimated as  ${E_{\rm{k}}} \simeq \frac{1}{{16}}M_{\rm PBH}{c^2}{l^{ - 1}}$. Accordingly, the corresponding temperature around merger is $T \simeq {E_{\rm{k}}}/{k_{\rm{B}}} \simeq {10^{63}}$K for $M_{\rm PBH} \sim 10^{28}$g, where ${k_{\rm{B}}}$ is the Boltzman constant. Therefore, the accelerating charged BHs have potential for radiating high brightness temperature GHz radiation seen in FRBs, as high as $>10^{37}$K \cite{Katz 2014}. This actually is the coherent emission by `bunching' of a large amount of charges in the hole i.e. $P \propto {q^2}$.

 For the duration of FRBs, as shown in Refs. \cite{Lorimer 2007,Thornton 2013}, also pointed out by \cite{Katz 2014}, most FRBs were temporally unresolved. And others like that of \cite{Lorimer 2007}  were temporally resolved but their widths are explained as multipath dispersion of travel times due to scattering. The similar discussion is also seen in \cite{Keane 2016}. Theoretically, the broadened pulse width can be estimated as $W \sim D_{L} \theta_{sc}^{2}/2c \sim (D_{L}/1 {\rm Gpc}) \times (10^{10} \theta_{sc} / 3)^{2} \times 5 ~ {\rm ms} $ in the context of thin screen approximation \cite{Williamson 1972}, where $D_{L}$ is luminosity distance of the source and $\theta_{sc}$ is the corresponding scattering angle, which may account for the observed millisecond pulses of FRBs \cite{Katz 2016, Xu 2016}. It indicates two important facts: 1) the observed duration of most FRBs are the upper limit of the intrinsic one  and 2) the pseudo-luminosity \cite{Keane 2012} correspond to the observed duration are only the lower limit of the intrinsic luminosity. This means that we can only get the lower limit of q if we imply Eq.\eqref{eq-P} to FRBs with pseudo-luminosity for most cases. However, the total energy of FRBs is unambiguous in the band of observation, which is more physical than the luminosity. Based on the points above, we will use the energy (i.e.~Eq.\eqref{eq-E}) to study the energetics of FRBs instead of luminosity in this work. The intrinsic time scale of radio burst in our model, according to Eq.\eqref{eq-tau}, can be estimated as $\tau \sim (M_{\rm PBH}/10^{28} {\rm g}) \times 1 {\rm ns}$, which is in order of nanoseconds and much smaller than millisecond. Apparently, this extremely short intrinsic duration can't be resolved by Parkes.

As discussed in previous paragraph, the intrinsic broad band spectrum of radio bursts is a power law with exponential cutoff. However, we may not obtain an FRB's intrinsic spectrum observationally due to its extremely short intrinsic duration. Instead, we could obtain the time-integrated spectrum by observation as ${{F}_{\nu }}\propto \text{d}E/\text{d}\nu \propto {{\nu }^{-1}}$ with exponential cutoff at $\nu_{\rm{p}}$ ($\nu {\rm{ = }}\omega /2\pi$), which may naturally explain the spectral indices of FRBs detected in narrow bandpass\cite{Deng 2018}. This is consistent with the recent works \cite{Macquart 2018} and \cite{Shannon 2018} which found the mean time-integrated spectral index of FRBs is $\alpha  =  - 1.6_{ - 0.2}^{ + 0.3}$ and $\alpha  =  - 1.8 \pm 0.3$ (${F_\nu } \propto {\nu ^\alpha }$) respectively. In conclusion, the major features of FRBs can be well explained by our model.

\section{MERGER RATE and ABUNDANCE of PBHs}
As discussed in the last section, we consider the energy rather than the luminosity of FRBs. Assuming that those observed non-repeating FRBs are originated as proposed in our model, one derives from Eq.\eqref{eq-E} that, for $q \sim 10^{-4.5} (E_{\rm FRB}/10^{40} {\rm erg})^{1/2}(M_{\rm PBH} / 10^{28} {\rm g})^{-1/2}$, our model may explain the FRBs with released energy at $O(10^{40}) {\rm erg}$ \cite{Deng 2018}. The parameter $q$ holding by the PBH binary is compatible with the results reported in Refs. \cite{Liebling 2016, Zhang 2016}, which attempted to explain the gamma-ray transient associated with GW150914 by charged BH binary coalescence. Hereafter, we use $q$ to label the FRBs instead of the energy $E$ by the transformation relation $E\simeq q^{2}M_{\rm PBH}c^{2}$ with $M_{\rm PBH}\simeq 10^{28} {\rm g}$. An FRB with $q$ means that this FRB's observed energy is $q^{2}M_{\rm PBH}c^{2}$. For the coalescence of PBH binaries, we define the local specific event rate density (local merger rate density per unit charges holding by the PBH binary) $n_{0,q}\equiv n_{0}\phi(q)$, where $\phi(q) \propto dN/dq$ is the probability distribution for $q$ of the PBH binaries normalized by $\int_{0}^{1} \phi(q)dq=1$. Therefore, the local event rate for coalescence of PBH binaries with charges larger than $q$ can be defined as $n_{0,>q}=n_{0}\int_{q}^{1} \phi(q')dq'$.

\begin{figure}[h]
\epsfig{figure=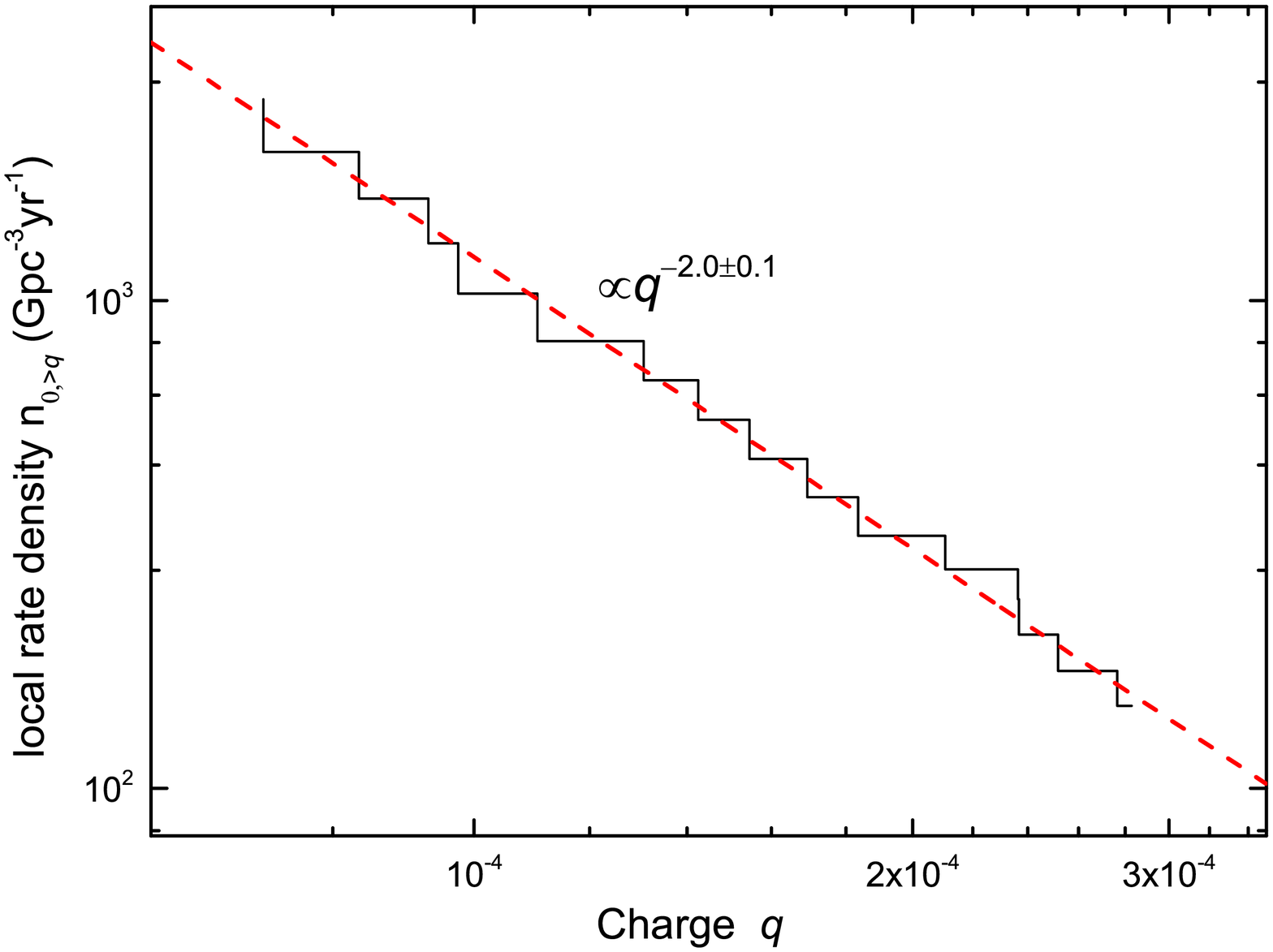,width=3.5truein}
\vskip-0.28in
\caption{The distribution of local event rate of FRBs with charges larger than the charge parameter $q$ by using the current sample of the observed FRBs, i.e., $n_{0,>q}$ as described in our model (depicted by the black solid line).
The red dash line is the best fit with a single power law to the data, resulting $n_{0,>q} \propto q^{-2.0\pm0.1}$.
It also implies that the charge distribution function evolves as $\phi(q)\propto q^{-3.0\pm0.1}$ for $q\geq 7.2\times10^{-5}$.
}
\label{fig:f1}
\end{figure}

To depict the amount of PBHs, one usually introduce the fraction of PBHs against the matter sector $f_{\rm PBH}\equiv \Omega_{\rm PBH}/\Omega_{M}$.
For the convenience of astrophysical consideration, we also introduce the dimensionless mass of PBHs as $m_{\rm PBH} \equiv M_{\rm PBH}/M_{\odot}$,
where $M_{\odot}$ is the mass of the Sun. The event rate density $n(z)$ for merger of PBH binary is calculated by  $n(z) = {n_0}{(1 + z)^3}{({t_0}/{t_{\rm{z}}})^{34/37}}$ \cite{Ali 2017}, where $n_{0}$ is the local merger rate as follows,
\begin{equation} \label{eq-n0}
 n_{0} \simeq 1.3 \times 10^{-3} \frac{\rho^{0}_{M}}{M_{\odot}t_{0}} m_{\rm PBH}^{-32/37} f_{\rm PBH}^{2} (f_{\rm PBH}^{2}+\sigma_{ \emph eq}^{2})^{-21/74} ~,
\end{equation}
$t_{0}$ and $t_{\rm z}=\int_{z}^{\infty} [(1+z')H(z')]^{-1} dz'$ is the age of the universe at present and redshift $z$, respectively. $\rho^{0}_{M}$ is the matter density of the universe at $t_{0}$, $\sigma_{eq}=5 \times 10^{-3}$.
Moreover, for a radio telescope survey with fluence limited sensitivity $F_{th}$, field of view $\Omega$, and operational time $T$, the observational number of FRBs within the range ($q, q+dq$) can be calculated by
\begin{equation} \label{eq-dN}
 dN{\rm{ = }}\frac{{\Omega T}}{{4\pi }}\phi (q)dq\int_0^{{z_{\max }}(q)} {\frac{{n(z)}}{{1 + z}}} \frac{{dV(z)}}{{dz}}dz ~,
 \end{equation}
in our model. Note that, in the above equation $T$s is the observational duration of each point and $\Omega$ is the sky area of the survey. For Parkes, we adopt $\Omega T = 144.1\,{\deg ^2}{\rm{h}}$ per burst as in \cite{Champion 2016}. Additionally, $dV(z)$ is the comoving volume element, $z_{\rm max}(q)$ is the maximum redshift where the FRBs arisen from the PBH binary coalescence with $q$ can be detected by a radio telescope with fluence limited sensitivity $F_{th}$. It is defined as
\begin{equation}
 {F_{th}} = \frac{(1+z_{\max })E}{{4\pi D_L^2({z_{\max }})}} \simeq \frac{{{(1+z_{\max })q^2}{M_{\rm PBH}}{c^2}}}{{4\pi D_L^2({z_{\max }})}} ~.
\end{equation}

We use a sample of $28$ FRBs presented in \cite{Deng 2018}, which were detected by the Parkes telescope. Accordingly, this sample is free from the issue caused by different telescope parameters. By using the same technique as in \cite{Sun 2015}, Eq. \eqref{eq-dN} can yield an estimate on the local event rate of FRBs above a certain $q_{i}$ in our model.
To be specific, it yields
\begin{equation}
 {n_{0, > {q_i}}} \simeq \frac{{4\pi }}{{\Omega T}}\sum\limits_{{q_i}}^{{q_{\max }}} {\frac{1}{{f({q_i})}}} ~,
\end{equation}
where is the maximum value of $q$ in the FRBs sample, and
\begin{equation}
f(q) = {\int_0^{{z_{\max }}(q)} {{{(1 + z)}^2}\left( {\frac{{{t_0}}}{{{t_{\rm{z}}}}}} \right)} ^{34/37}}\frac{{dV(z)}}{{dz}}dz.
 \end{equation}
For Parkes, $F_{\rm th} = 5.4 \times 10^{-18} {\rm erg}/{\rm cm}^{2}$ \cite{Deng 2018}, and hence, we obtain the local event rate $n_{0, >\rm q}$ as function of $q$, $n_{0, >\rm q} \propto q^{-2.0\pm0.1}$, as shown in Fig. \ref{fig:f1}. It also implies that the charge distribution takes the form of $\phi(q)\propto q^{-3.0\pm0.1}$, which is a power law function with a steep index $-3.0$ for $q\geq q_{\rm min}$, where $q_{\rm min}=7.2\times 10^{-5}$ is the minimum of $q$ in this sample. For $q=q_{\rm min}$, we get the local event rate with $q>q_{\rm min}$
\begin{equation} \label{eq-n0qmin}
 {n_{0, > {q_{\min }}}} \simeq 2.6 \times {10^3}\left( {\frac{{144.1{{\deg }^2}{\rm{h}}}}{{\Omega T/N}}} \right){\rm{Gp}}{{\rm{c}}^{ - 3}}{\rm{y}}{{\rm{r}}^{ - 1}} ~.
 \end{equation}
Combining Eqs. \eqref{eq-n0} and \eqref{eq-n0qmin}, one gets the inequality $n_{0}\geq n_{0,>q_{\rm min}}$, which further implies
\begin{equation}
 {{f}_{\rm PBH}} \gtrsim {{10}^{-5}} ~~ \text{for} ~~ {{M}_{\rm PBH}}\sim {{10}^{28}} {\rm g} ~.
\end{equation}
Here we get a lower bound of $f_{\rm PBH} \gtrsim 10^{-5}$. Accordingly, if one expects that the coalescence of PBHs satisfying a certain charge distribution as described by the present model are responsible for the observed non-repeating FRB event, then, the fraction of PBHs contribution to the matter sector in the Universe should $> O(10^{-5})$. It is interesting to note that, the current upper limit on $f_{\rm PBH}$ at $M_{\rm PBH}\sim 10^{28}$ g is estimated as $\sim O(0.1)$ as given by microlensing observation \cite{Tisserand 2007, Mediavilla 2009, Carr 2016}.

\section{Conclusion and Discussion}
In this paper, we propose the radio bursts from PBHs binaries coalescence as an alternative model for non-repeating FRBs. The FRBs' radiated band, spectrum, released energy and event rate can be well explained by the model. With the current sample of FRBs, we obtain the charge distribution $\phi(q)\propto q^{-4.5}$ of PBHs for $q\geq q_{\rm min}$. With the hypothesis that FRBs in the current sample are originated as proposed in our model, we can get a lower bound $f_{\rm PBH}\gtrsim 10^{-5}$ for the abundance of the PBHs with  $M_{\rm PBH}\sim10^{28}$g needed to explain the observed rate of FRBs. The further observation is needed to identify the physical origin of FRBs to make ideal constrains to the abundance and the charges of PBHs. A charged black hole has not yet been discovered in the past. In the future, the gravitational waves produced by mergers of small black holes can be detected by high frequency gravitational wave detectors \cite{Cruise 2000, Ballantini 2003, Baker 2006, Nishizawa 2008, Li 2009}. Whether or not there are electromagnetic signals associated with those gravitational waves, we can get important clue for charged PBHs.

As so far there is no concrete evidence for physical origin of FRBs, it is expected that the future FRB observations would provide more information that can be applied to falsify various models include the present one. Correspondingly, it deserves to point out a distinguishable features derived in our model that would be of observable interest, i.e., the FRBs originated from our model could carry the dipole-quadrupole feature in their radiation spectrum. If this particular pattern were detected in the future FRB experiments, it could be a smoking gun for the verification of the physical origin of FRBs from charged BH coalescence and put an extremely tight constraint on the PBH fraction.

Finally, we must say that attaining and maintaining the $q \sim 10^{-4.5}$ for producing an FRB in this model seems small but it is actually quite large. Indeed, Reissner-Nordstr$\ddot{\rm o}$m BHs have charge greater than $q \simeq \sqrt G {M_{\rm{p}}}/e \simeq {10^{ - 18}}$ would cause charges separation in the surrounding plasma and decrease the q by attracting the opposite charges, where $M_{{p}}$ and $e$ is the mass and charge of the proton respectively. Nevertheless, Wald showed that a Kerr BH in the magnetic field $B$ would be charged up to \cite{Wald 1974}
\begin{equation}
q \simeq {10^{ - 4.4}}\,\,a\left( {\frac{{{M_{{\rm{BH}}}}}}{{{{10}^{28}}g}}} \right)\left( {\frac{B}{{{{10}^{20}}{\rm{G}}}}} \right),
\end{equation}
where $a = J/M_{BH}^2$ is the Kerr parameter. For PBHs, the temperature of the universe is $T \simeq {10^{15}}{\rm{K}}\,{(t/{10^{ - 10}}{\rm{s}})^{{\rm{ - }}1/2}}$ \cite{Weinberg 1972} at the forming time-scale $t\sim 10^{-10}$s with mass $\sim 10^{28}$g. Assume a fraction $\eta $ of the energy transforming into magnetic energy ${B^2}/8\pi  \sim \eta {a_{\rm{u}}}{T^4}$, where ${a_{\rm{u}}}$ is the energy density constant. Accordingly, we have $B \sim {10^{22}}{\rm{G}}\,\,\eta _{0.01}^{1/2}{(t/{10^{ - 10}}{\rm{s}})^{ - 1}}$. Therefore, there are strong enough magnetic fields in the extremely early universe after PBHs born. On the other hand, the holes would spin up by accreting surrounding dense plasma. This shows that there is a chance for PBHs significantly charged up in the early universe, within the charging time-scale ${t_{\rm{c}}} \sim {r_s}/c \simeq {10^{ - 10}}{\rm{s}}$. The question is how the PBHs can keep the charges until merger.  As discussed in \cite{Zhang 2016}, the charges may be able to maintain stably by the co-rotating magnetosphere formed around the charged holes similar to pulsars \cite{Blandford 1977,Michel 1982}. However, the maintaining time-scales is difficult to know.

There is a more promising way to charge PBHs, as pointed out in Ref.\cite{Liebling 2016}, which involves magnetic monopoles. And the electrodynamics of PBHs with magnetic monopoles is the same as discussing in our model due to the symmetry between electric and magnetic charges.  Magnetic monopoles produced abundantly in the extremely early universe are a generic prediction of grand unified theories, with mass ${M_{\rm{m}}} \simeq {10^{ - 8}}{\rm{g}}$ and magnetic charge ${g_{\rm{m}}} \simeq 3.3 \times {10^{ - 8}}{\rm{e}}{\rm{.s}}{\rm{.u}}{\rm{.}}$ \cite{Preskill 1984}. \cite{Stojkovic 2005} shows that PBHs can sufficiently accrete magnetic monopoles in the early universe then avoiding the so called monopole problem. Assume that the maximum charge $Q_{m}$ accreted by the hole achieves when the the magnetic force balances gravitational force  i.e.
\begin{equation}
\frac{{G{M_{{\rm{PBH}}}}{M_{\rm{m}}}}}{{{r^2}}} \simeq \frac{{{Q_{\rm{m}}}{g_{\rm{m}}}}}{{{r^2}}}.
\end{equation}
Accordingly, we have ${q_{\rm{m}}} = {Q_{\rm{m}}}/\sqrt G {M_{{\rm{PBH}}}} \simeq \sqrt G {M_{\rm{m}}}/{g_{\rm{m}}} \simeq {10^{ - 4.1}}$. Interestingly, this value is very close to the charges needed to produce FRBs in our model. In addition, the magnetic charges of PBHs are difficult to be neutralized unlike the electric charges. Therefore, PBHs can survive the long evolutionary history of the entire universe to keep the magnetic charges before merger. It's fascinating that perhaps the FRBs really are the signals of PBHs and magnetic monopoles.

\section*{Acknowledgments}
We thank J.-J. Wei, B. Yu and P. Zhang for helpful discussion. This work is supported by the National Basic Research Program (¡°973¡± Program) of China (grant No. 2014CB845800) and the National Natural Science Foundation of China (grant Nos.  11433009, 11533003, 11653002, 11673068, 11722327, 11725314 and 11851304).
YC is supported in part by China's Youth Thousand Talents Program, by CAST Young Elite Scientists Sponsorship Program (2016QNRC001), and by the Fundamental Research Funds for the Central Universities. X.F.W is also partially supported by the Youth Innovation Promotion Association (2011231), the Key Research Program of Frontier Sciences (QYZDB-SSW-SYS005) and the Strategic Priority Research Program ¡°Multi-waveband gravitational wave Universe¡± (grant No. XDB23040000) of the Chinese Academy of Sciences. E.W.L also acknowledges support by the Guangxi Science Foundation (grant No. AD17129006).


\end{document}